\documentclass{epsconf}
\usepackage{graphicx}
\usepackage{wrapfig}
\usepackage{amsmath}
\pdfoutput=1
\title{Phase dependent advection-diffusion in drift wave - zonal flow turbulence}
\author{\underline{Sara Moradi}$^1$ and Johan Anderson$^2$}
\institute{$^1$ Fluid and Plasma Dynamics, Universit\'{e} Libre de Bruxelles, 1050-Brussels, Belgium \\
$^2$ Chalmers University of Technology, SE-412 96 G\"{o}teborg, Sweden}

\begin{document}
\maketitle
In plasma turbulence theory, due to the complexity of the system with many non-linearly interacting waves, the dynamics of the phases is often disregarded and the so-called random-phase approximation (RPA) is used assuming the existence of a Chirikov-like criterion for the onset of wave stochasticity \cite{Zakharov1984}. The dynamical amplitudes are represented as complex numbers, $\psi=\psi_r+i\psi_i=ae^{i\theta}$, with the amplitudes slowly varying whereas the phases are rapidly varying and, in particular, distributed uniformly over the interval $[0;2\pi)$. However, one could expect that the phase dynamics can play a role in the self-organisation and the formation of coherent structures as was shown in ref. \cite{GuoPRL2015}. In the same manner it is also expected that the RPA falls short to take coherent interaction between phases into account. In this work therefore, we studied the role of phase dynamics and the coupling of phases between different modes on the characteristic time evolution of the turbulent. We assume a simple turbulent system where the so-called stochastic oscillator model can be employed. The idea of interpreting turbulence by stochastic oscillators goes back to Kraichnan \cite{Kraichnan1961}. The stochastic oscillator models can be derived from radical simplifications of the nonlinear terms in the Navier-Stokes or Gyro-Kinetic equations. In this particular case we adopt the basic equation for the stochastic oscillator model with passive advection and random forcing from Ref. \cite{krommes2000b}:
\vspace{-2mm} 
\begin{eqnarray}
\partial_t \psi + u(t).\nabla \psi = \hat{f}^{ext}(t),
\label{eq0}
\vspace{-2mm}
\end{eqnarray} 
where $u(t)$ and $\hat{f}^{ext}(t)$ are random values with given statistical properties. Following the work performed by Krommes in Ref. \cite{krommes2000b} on the impact of random flows on the fluctuation levels in simple stochastic models, we consider the passively advected fluctuations of a scalar $\psi$ such as temperature, to obey
\begin{eqnarray}
\vspace{-3mm}
&&\partial_t \delta\psi(\mathbf{x},t) + \delta \mathbf{V}(\mathbf{x},t). \nabla \delta\psi - D \nabla^2 \delta\psi = \delta f (\mathbf{x},t)
\label{eq1}
\vspace{-3mm}
\end{eqnarray}

As in Ref. \cite{krommes2000b} to keep the discussion as general as possible the linear physics is modelled by a random external forcing $\delta f$ and a classical dissipation $D \nabla^{2}$. Furthermore, we assume homogeneous statistics and thus only solve eq. (\ref{eq1}) in one dimension, i.e. $ \delta \psi(y,t)$. Here, $\delta \mathbf{V}$ is a statistically specified random flow velocity, corresponding to the random $u(t)$ in eq. (\ref{eq0}). In the present model we ignore the spatial dependence of $\delta \mathbf{V} (\mathbf{x},t)= \delta \mathbf{V}(t)$ employing a Fourier transformation of eq. (\ref{eq1}) we obtain the forced stochastic oscillator equation (see eq. (32) in Ref. \cite{krommes2000b})
\vspace{-4mm}
\begin{eqnarray}
\vspace{-2mm}
&&\partial_t \delta\psi_k + i \delta u_k(t) \delta\psi_k + \nu_{k} \delta\psi_k = \delta f_k(t)
\label{Feq1}
\vspace{-2mm}
\end{eqnarray}
where we assumed the following
\vspace{-4mm}\begin{eqnarray}
&& \delta u_k(t) = k_y \bar{V}\exp(i\theta_k(t))\label{eq4}\\
&& \delta f_k(t) = \gamma \exp(i\phi_k(t))\label{eq5}
\vspace{-4mm}
\end{eqnarray}

In our model, thus, the random flow and forcing are assumed to be similar to oscillators with constant amplitude and phases that varies in time with $\theta_k(t)$ and $\phi_k(t)$, respectively. Here, we used the same definitions for $\gamma\dot= 2\kappa^2 D$, and $\nu_k = k^2 D$ as in Ref. \cite{krommes2000b} with $\kappa$ being the a constant measuring the strength of the forcing. $\bar{V}$ is a constant measuring the strength of the random flow. The scale dependence is introduced by the multiplication by $k_y$ and the prescribed dispersion relation for the natural frequency, see eq. (\ref{eq6}) in the next section. In general one can assume also a stochastic amplitude and therefore treat $\bar{V}$ as a random value with a given statistical property, which would allow for further degrees of freedom in the model. However at this point we aim to study the role of phase self-organisation on the fluctuating scalar and therefore we will assume a constant amplitude.

\section{Phase coupling model}
The dynamics of the phases are described by the two coupled first order differential equations as
\vspace{-6mm}\begin{eqnarray}
&&\dot{\theta}_{k}(t)=\omega_k+(2\pi)^{-1}\sum_{i=1}^{N}J_{ik}sin(\theta_i-\theta_k)+\frac{1}{2}\epsilon\phi_{k},
\label{theta1}\\
&&\dot{\phi}_{k}(t)=\zeta_k+(2\pi)^{-1}\sum_{i=1}^{N}S_{ik}sin(\phi_i-\phi_k)-\frac{1}{2}\epsilon\theta_{k},
\label{theta2} \;\;\;\;\;\;(k=1,...,N).
\vspace{-6mm}
\end{eqnarray}
where the $\theta_k$ and $\phi_k$ follow a non-linear sinusoidal coupling as in the Kuramoto model \cite{kuramoto} with additional linear cross coupling between the two phases. Here, the analogy is that $\phi$ is the phase corresponding to a fluctuating radial excursion associated with the drift wave, and the $\theta$ is the phase corresponding to the oscillating zonal shear modulating the direction of excursion via eddy tilting \cite{moradi2Dmodel, moradi}. The linear cross-coupling introduces a cross correlation between the two motions similar to the regular Lotka-Volterra predator-prey model. 
This cross-coupling mimics the interactions between the drift wave turbulence and zonal flows which have been observed to follow self-consistent feedback loop systems similar to predator-prey trends. Here, $\dot{\theta}_{k}(t)$, and $\dot{\phi}_{k}(t)$ denote the time derivatives of the phases of $k$th mode.

\begin{wrapfigure}{r}{70mm}\centering
\vspace{0mm} 
\includegraphics[width=4cm, height=3cm]{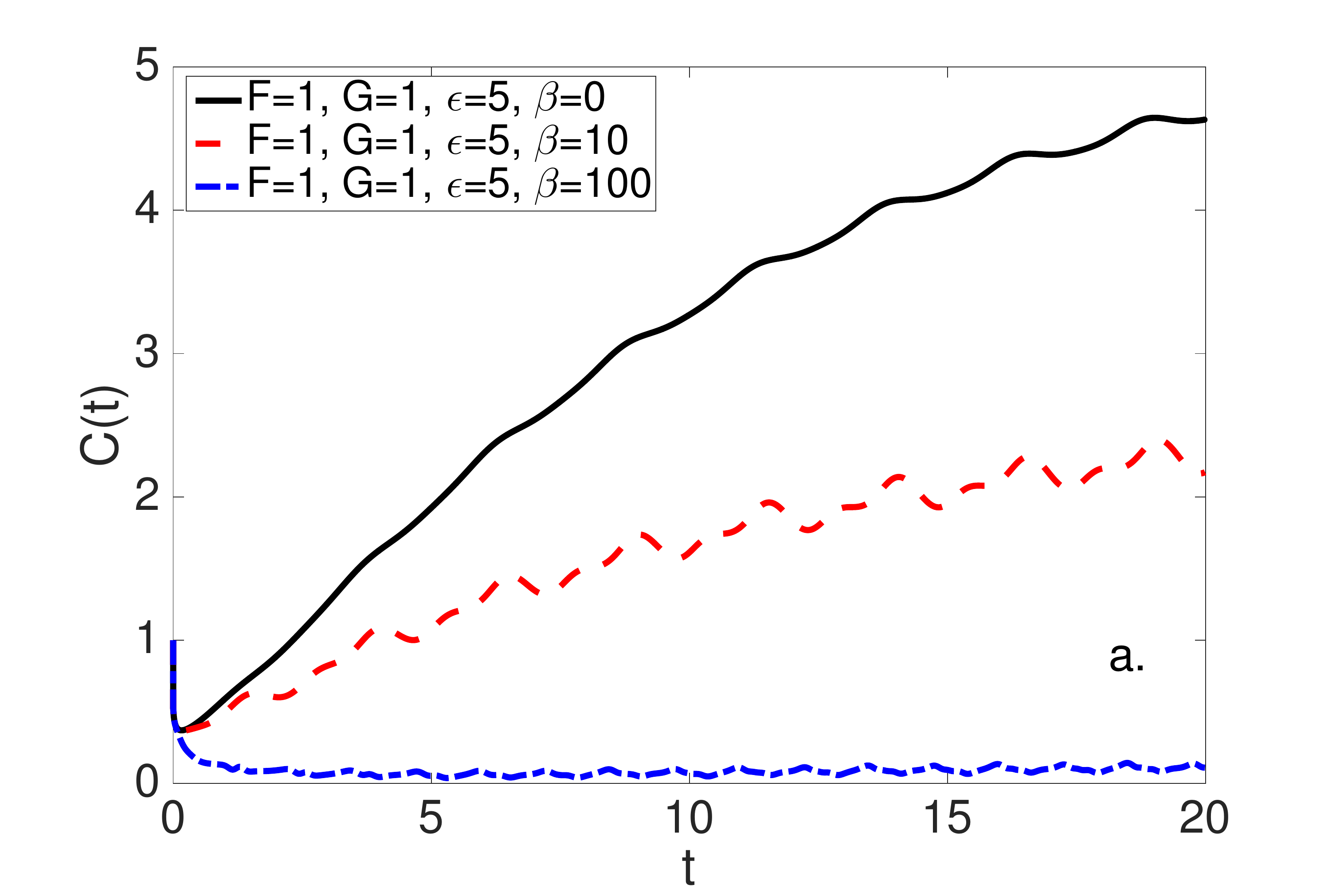}\includegraphics[width=4cm, height=3cm]{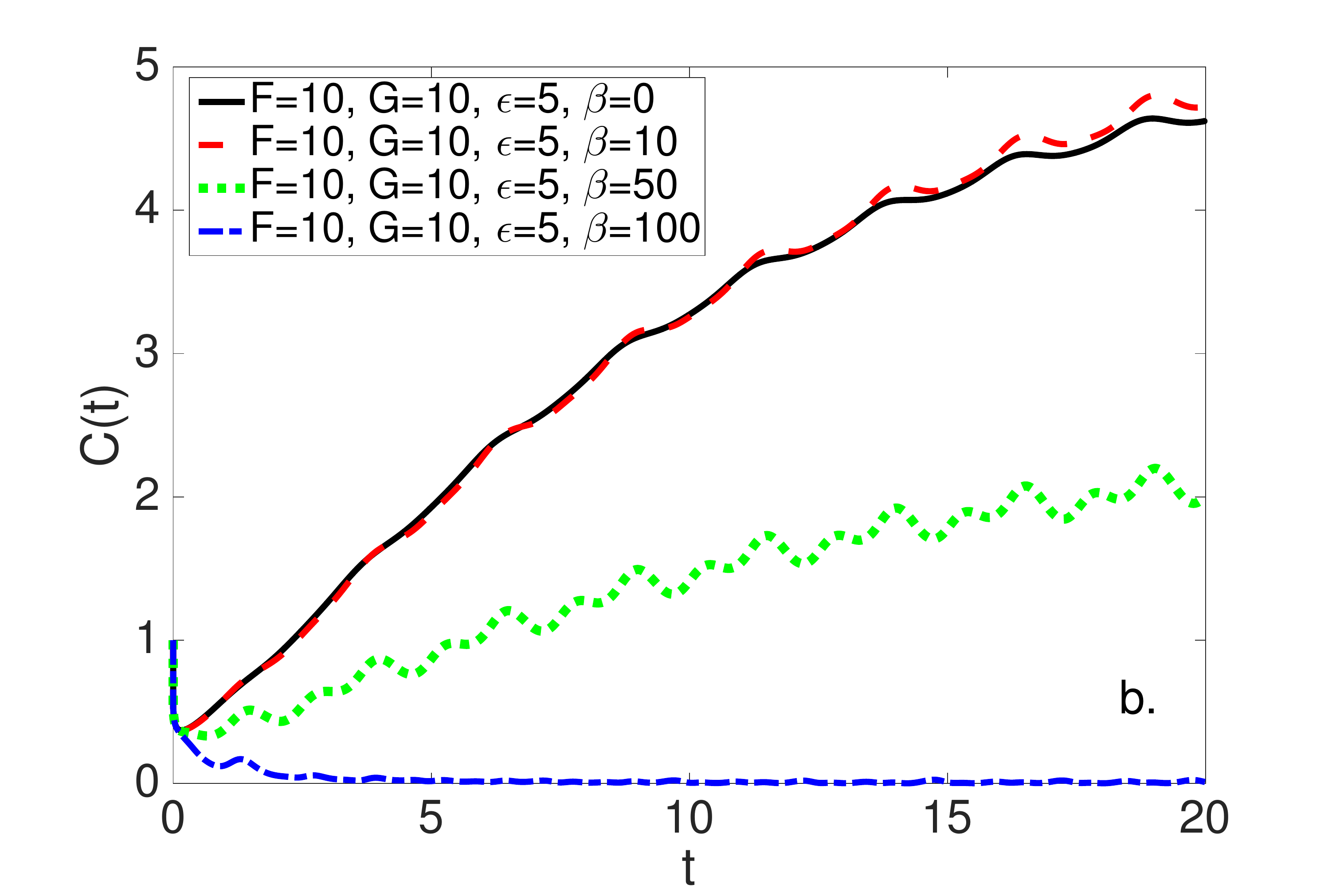}
\caption{\label{fig10} (a) Time evolution of sample averaged $C(t)$ for different $\beta$ at $\epsilon=5$ and (a) $F=G=1$ (b) $F=1, G=10$. (Solid lines) show the results for $\beta=0$, (dashed lines) $\beta=10$, and (dashed-dotted lines) $\beta=100$. In (b) (green dotted line) shows the case with $\beta= 50$.}
\vspace{-3mm} 
\end{wrapfigure}

The parameter $\epsilon$ is a free parameter and modifies the strength of the linear cross-coupling term and $N$ is the number of modes considered. $\omega_k$ are the natural frequencies of the flow assumed randomly distributed according to a Gaussian distribution with zero mean, $f(\omega)=\exp(-\omega^2/2)/\sqrt 2\pi$, and $\zeta_k$ are the natural frequencies of the forcing, also assumed randomly Gaussian distributed, where the mean is defined through a dispersion relation similar to that of the DWs (see Ref. \cite{GurcanPRL2009}):
\vspace{-2mm}
\begin{eqnarray}
&&<\zeta>(k) = -\beta\frac{k}{1+k^2}\label{eq6}
\vspace{-2mm}
\end{eqnarray}
where $\beta$ is a free parameter of the model. In the usual DW picture this parameter represents a gradient e.g. density gradient, $\delta n/\delta y$.
$J_{ik}$ and $S_{ik}$ measure the strength of the interactions between oscillator $i$ and $k$ in each population, and they are assumed randomly Gaussian distributed with standard deviation defined as $\sigma_{\theta,\phi}=\{F,G\}/(\sqrt(2N))$. Note, that low values $F,\;G$ correspond to weak coupling while high values correspond to strong coupling.

\section{The numerical set up}
The numerical integration of eq. (\ref{Feq1}) is performed using the Runge-Kutta 4th order scheme (RK4) the sampling time step is $\Delta t=0.01$. For initial conditions we use $|\delta\psi_k(0)|=1$, with the phases of $\delta\psi_k(0)$ set to zero. For the random flow and the forcing we set the initial phases as $\theta_k(0)=\phi_k(0)=0$. The mode number $k$ is chosen following a shell model by setting $k_n = k_0 \times g^{n}$ where $n=1,\dots N/5$ where $N=125$ corresponds to the number of modes, with $k_0=1$ and $g=1.25$. An averaging over $N_{s}=10$ samples of $J_{ik}, S_{ik}$ is also performed. 

\section{Results of numerical simulations} 
\begin{wrapfigure}{r}{70mm}\centering
\vspace{0cm} 
\includegraphics[width=6cm, height=8cm]{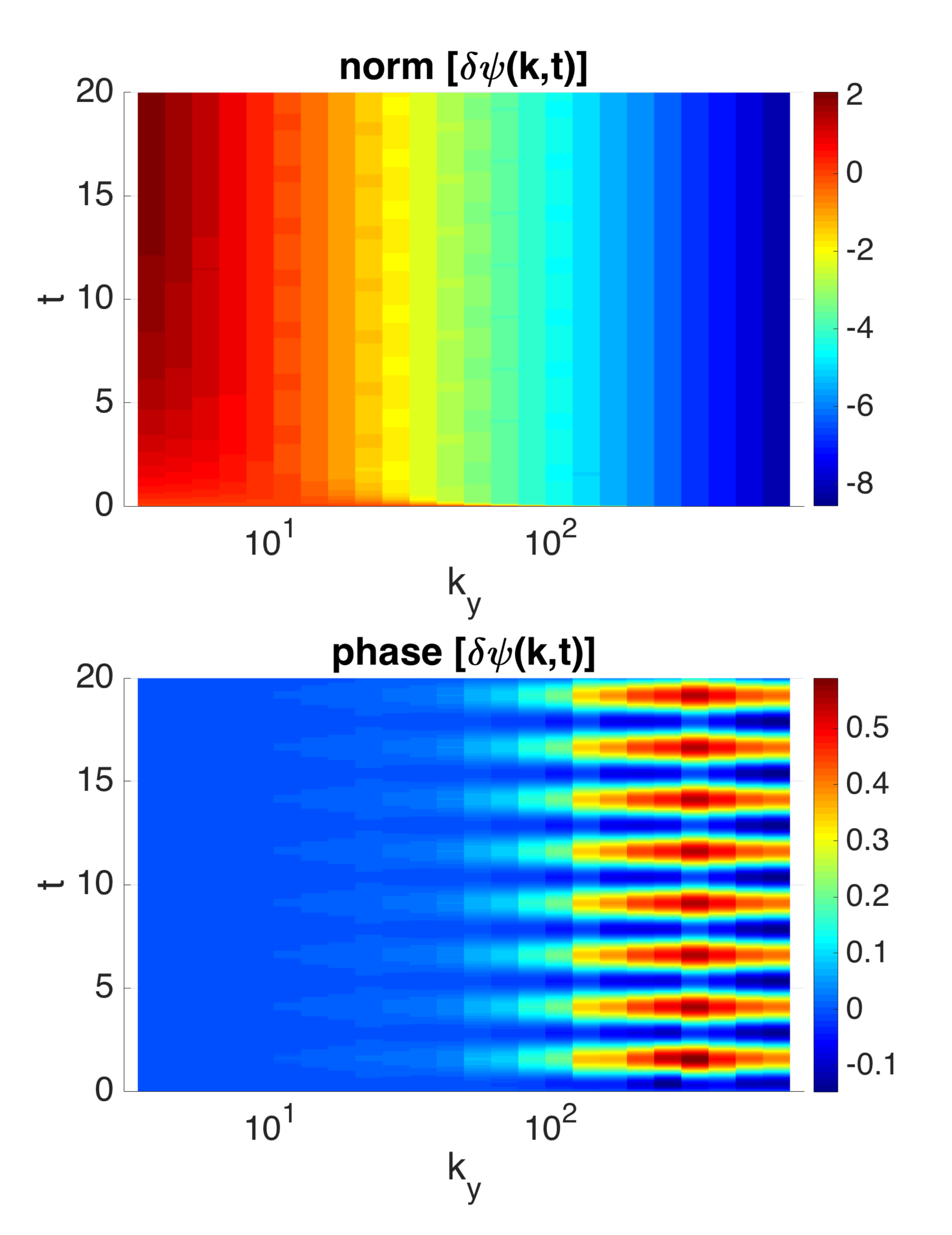}
\caption{\label{fig11} (top) The logarithm of norm and (bottom) the phase of $\delta\psi_k$ as functions of mode number $k$, and time $t$. Here, $\beta=100$, $\epsilon=0$, and $F=G=1$. Similar results are found for the case with $F=G=10$.}
\vspace{-3mm} 
\end{wrapfigure}

Figures \ref{fig10} (a and b) illustrate the time evolution of $C(t)$ for different phase states and increasing values of $\beta$. When $\beta =0$ the fluctuation auto-correlation increases in time similarly in both synchronised and a-synchronised cases i.e. $F=G=1$, and $F=G=10$, respectively.The time evolution of the norm and phase of $\delta\psi_k$ are shown in Figs. \ref{fig11}. Here we observe a separation between the high and low $k$ where in the low $k$, the phases sit at $0$ while the phases at high $k$ oscillate in an small angle between $0$ and $0.5$ with regular frequency. The increase in $\beta$ results in a strong reaction of the saturation levels in both $F=G=1$, and $F=G=10$ states, and as the increase in $\beta$ desynchronises the phases of the forcing the saturation levels converge to the same value for both states. Small modulations are observed in the evolution of $C(t)$ when $F=G=10$ which decreases in amplitude as $\beta$ is increased.

The effect of high $\beta$ on the evolution of the phases and norms, is to increase the frequency and the range of oscillation of the phases in a wide range in $k$, while the norm is strongly affected at low $k$ as it decays in time. 
 
In this work, we have introduced a non-linear phase coupling model into the simplest model of the passive advection-diffusion of a scalar with forcing. The phase-coupling follows the well-established Kuramoto paradigm that has been shown to represent systems displaying self-organisation well. Our results show that the assumption of a fully stochastic phase state of the turbulence is more relevant for high values of scale separation with the energy spectrum following a $k^{-7/2}$ decay rate, while for lower scale dependence the a-synchronised and synchronised phases differ significantly, and one could expect the formation of coherent modulations in the latter case.

\end{document}